\documentclass[11pt,twoside]{article}

\usepackage{asp2006}
\usepackage{epsf}
\usepackage{graphicx}
\usepackage{lscape}
\usepackage{amsbsy}
\usepackage{amssymb}

\begin{document}
\title{2D radiative modelling of He\,{\sc i} spectral lines formed in solar
prominences}

\author{Ludovick L\'eger and Fr\'ed\'eric Paletou}   

\affil{Universit\'e de Toulouse, Laboratoire d'Astrophysique de
Toulouse-Tarbes, CNRS, 14 ave. E. Belin, F-31400 Toulouse, France}

\begin{abstract} 
We present preliminary results of 2D radiative modelling of He\,{\sc
i} lines in solar prominences, using a new numerical code developed by
us \citep{leger07}. It treats self-consistently the radiation transfer
and the non-LTE statistical equilibrium of H and, in a second stage,
the one of He using a detailed atomic model. Preliminary comparisons
with new visible plus near-infrared observations made at high spectral
resolution with TH\'eMIS are very satisfactory.
\end{abstract}

\section{Introduction}   
Multi-dimensional radiative transfer is of general interest in various
domains of astrophysics. Hereafter our primary interest concerns the
radiative modelling of isolated and illuminated structures in which
non-LTE plasma conditions prevail: solar prominences. In particular,
spectral lines of He\,{\sc i} are commonly used for the diagnosis of
magnetic fields in solar prominences. While most inversion codes,
except {\sc hazel} \citep{hazel}, assume that spectral lines such as
$D_{3}$ and 1083 nm for instance are optically thin, it is not always
confirmed observationally \citep[e.g.][]{lopez02}. We also have a good
idea about conditions under which geometry effects may have an impact
on the formation of moderately thick lines \citep{pal97}. In order to
improve our ability to diagnose prominences using such spectroscopic
data, we have thus developed a new 2D numerical radiative transfer
code \citep{leger07}.  Hereafter we present the numerical strategies
we adopted and comment on the performances of this code. Finally,
preliminary results of the 2D radiative modelling of He\,{\sc i} lines
and a first comparison with new observations are shown.

\section{Context and state-of-the-art}
Solar prominences are dense and cool structures hanging in a hot and
low density corona \citep{eth}. They are often the source of CMEs, and
the magnetic field very likely plays a major role in the triggering of
those instabilities leading to the onset of such large solar plasma
ejections \citep[see for instance][and references therein]{spirosjcv}.

\begin{figure}[!ht]
\centering
\includegraphics[height=6 cm]{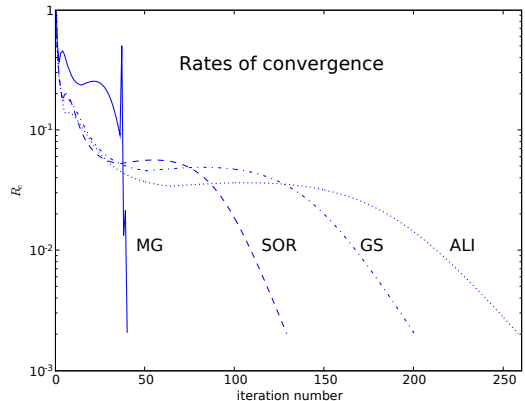}
\caption{Maximum relative change $R_{c}$ on the populations from an
iteration to another against the number of iterations for the MG, SOR,
Gauss-Seidel (GS) and ALI numerical schemes. MG provides a
considerable gain in number of iterations and computing time.}
\label{fig1}
\end{figure}

He\,{\sc i} lines such as $D_{3}$ and 1083 nm are among the best tools
for the study of magnetic fields in solar prominences. Full-Stokes
observations of these spectral lines, made at TH\'eMIS \citep{pal01}
have also led to a recent revision of magnetic field inversion tools
\citep{lopez02}.

The most recent radiative models \citep{labro01,labro04} assume 1D
static slabs and no atomic fine structure for the He model-atom,
leading to non-realistic gaussian synthetic triplet lines
profiles. However this fine structure is conspicuous in our
observations, and measurement of the ratio between the red and blue
peaks of $D_{3}$ and 1083 nm are often in contradiction with the
commonly assumed hypothesis of optically thin spectral lines
\citep[see][]{lopez02}.

It is therefore important to use the best numerical radiative transfer
methods, in 2D geometry at least, and a detailed He\,{\sc i} atomic
model in order to improve our ability: \emph{i)}
to compute realistic
synthetic spectral line profiles and then \emph{ii)} to improve
further available magnetic field inversion tools.

\section{Numerical strategies}

We adopted the following numerical strategies. The formal solution of
the radiative transfer equation is solved in 2D using the short
characteristics method of \citet{auer94}. To improve convergence, we
used a Gauss-Seidel iterative scheme with Successive Over-Relaxation
GS/SOR for multilevel atoms \citep{tf1,pal07}. Finally we included this
2D-GS/SOR iterative scheme into a nested multigrid method
\citep[see][]{fab97,leger07}.

Then, in order to synthesize He line profiles, we proceed as follows.
We solve the statistical equilibrium and the radiative transfer
equations for the hydrogen atom. Then the statistical equilibrium
equations for the helium atom are solved consistently with the
ionization equilibrium already computed for hydrogen.  For the first
step, we considered a plasma composed of neutral and ionized hydrogen,
and neutral helium with an abundance by number fixed at $0.1$.

As an illustrative example, we adopted a temperature $T =
8000\,\mathrm{K}$, a gas pressure $p_{g} =
0.05\,\mathrm{dyn}\,\mathrm{cm}^{-2}$ and a microturbulent velocity
$\xi = 5\,\mathrm{km}\,\mathrm{s}^{-1}$. The geometrical width of the
slab was $D_{y} = 5\,000\,\mathrm{km}$ and its height $D_{z} =
30\,000\,\mathrm{km}$, sampled with a 243x243 points spatial grid. We
considered just 5 bound levels and the continuum for hydrogen.

\begin{table}[!t]
\caption{Computing time and number of iterations (between parenthesis)
for four different numerical methods and two spatial grids.
\label{tab1}}
\begin{center}
{\small
\begin{tabular}{c c c c c}
\tableline\tableline
\noalign{\smallskip}
Grid size & MALI & GSM & SOR & MG \\
\noalign{\smallskip}
\tableline
\noalign{\smallskip}
163x163 & 184{\small mn}29{\small s} (156) & 120{\small mn}27{\small s} (123) & 78{\small mn}41{\small s} (84) & 31{\small mn}46{\small s} (29) \\
\hline
243x243 & 698{\small mn}38{\small s} (259) & 417{\small mn}24{\small s} (201) & 269{\small mn}21{\small s} (130) & 65{\small mn}31{\small s} (41) \\
\noalign{\smallskip}
\tableline
\end{tabular}
}
\end{center}
\end{table}

The respective rates of convergence for the ALI, Gauss-Seidel (GS),
SOR and multigrid (MG) multilevel iterative processes in 2D geometry
are displayed in Fig.~\ref{fig1} where it is plotted the maximum
relative change $R_{c}$ on the populations from an iteration to
another, against the number of iterations. In Tab.~\ref{tab1}, we also
provide the corresponding computing time and number of iterations for
two different spatial grids.

This shows that the MG scheme is definitely superior, both in
iteration numbers \emph{and} computing time, against all other
numerical schemes. This is especially true when the grid is more and
more refined. And we ought to remind again here that \emph{grid
refinement is very important for the sake of precision}, as
it was demonstrated by \citet{che03}.

\section{2D synthetic line profiles vs. observations.}

We anticipate that more realistic models should also consider the
multi-thread or \emph{spatial} fine structure, as observed in
prominences \citep[see][and references therein]{ph2007}.

Hereafter, we consider a single thread whose geometrical width and
height are taken identical, $D_{y,z} = 1\,200\,\mathrm{km}$, and
sampled with a 123x123 grid.  We also adopted five different
temperatures $T=$10\,000, 12\,000, 15\,000, 16\,000 and 17\,000 K, a
gas pressure $p_{g} = 1.0\, \mathrm{dyn}\, \mathrm{cm}^{-2}$ and a
microturbulent velocity $\xi = 5\,\mathrm{km}\,\mathrm{s}^{-1}$.

Concerning the atomic models, we considered 5 bound levels and the
continuum for hydrogen, but 17 bound levels (up to $n$=3) and the
continuum for helium, including the atomic fine structure for levels
$2^{3}P$, $3^{3}P$ and $3^{3}D$. We took explicitly into account the
hydrogen pumping for UV lines and the continuum of He, and the coronal
illumination for UV lines of He \citep{tob91}.

\begin{figure*}[!ht]
\centering
\includegraphics[height=7 cm]{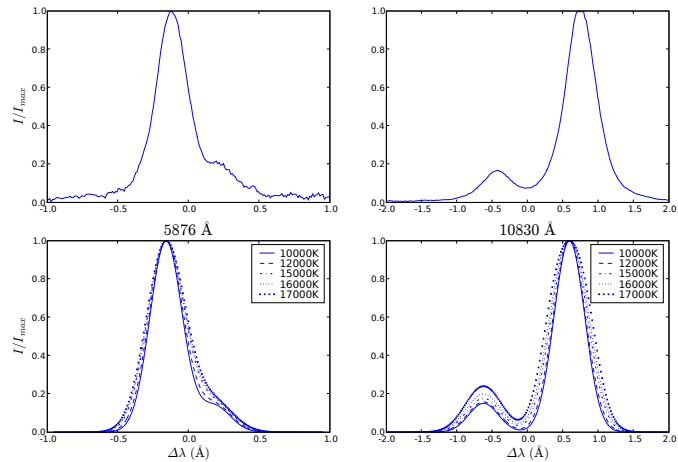}
\caption{Normalized (a) $D_{3}$ and (b) 1083 nm spectral lines
observed at TH\'eMIS in June 2007. Theoretical normalized line
profiles for (c) $D_{3}$ and (d) 1083 nm for 5 different temperatures
are displayed for comparison.}
\label{fig2}
\end{figure*}

Synthetic profiles and a sample of our observations made at TH\'eMIS in
June 2007 are plotted in Fig.~\ref{fig2}. At first glance, this
preliminary comparison with observations is very promising since we
are able to recover very nicely the high-resolution features of the
line profiles.

\section{Conclusions}
We have briefly presented preliminary results of the 2D radiative
modelling of He\,{\sc i} lines in solar prominences, using a new 2D
numerical radiative transfer code developed by us. So far, we favour
2D \emph{multi-thread} models in order to explain a number of
properties of observed He\,{\sc i} spectral lines. A more detailed
comparison with our new observations made at TH\'eMIS, simultaneously
in the visible and the near-IR, has begun. A next step will be to
evaluate the impact of such a 2D radiative modelling on magnetic field
inversion tools.

\end{document}